\documentclass[a4paper,10pt]{article}
\usepackage{graphicx}
\usepackage{longtable}
\title{Proposal for Compositeness of String out of Objects --
Fake Scattering, Finite String Field Theory Formulation 
\footnote{{\bf  23rd Bled Workshop 2020 ``What comes beyond the standard 
models'' 4-21 July 2020 Beld Slovenia}} }
\author{Holger Bech  Nielsen\footnote{Speaker}\\ Niels Bohr Institute, Copenhagen University, \\
E-mail: hbech@nbi.dk\\
\\
Masao Ninomiya\\
Yukawa Institute for Theoretical Physics,\\ Kyoto University, Kyoto 606-0105, japan\\ and\\
Yuyi Sugawara Lab., Science and Engineering,\\
 Department of Physics Sciences, Ritumeikan university\\
E-mail: msninomiya@gmail.com }
\date{Bled , July  , 2020}

\begin{document}
\maketitle
\begin{abstract}
We review a bit our earlier novel string field theory\cite{self2,self8} 
stressing 
the interesting property, that it becomes expressed in terms 
of particle like objects called by us ``objects'' which in our 
formalism do not at all develop in time. So in this way there is 
in our picture, in spite of it being supposed to reproduce string theory 
with an arbitrary number strings present - in this sense a string field 
theory -, in fact no time! This strange missing of time in the formalism 
gives rise to slight speculations about the philosophy of the 
concept of time. There is course then also no need for a Hamiltonian, but 
we construct or rather attempt to do so, a fake Hamiltonian or phantasy 
Hamiltonian,

\end{abstract}

{\bf Novel String Field Theory and Unitarity, although in non-relativistic 
case (Fake Scattering, Hope for finiteness).}
\section{Introduction}
\label{sec1} 
We have long worked on new/novel formulation \cite{self2} of the bosonic string 
theory 
as a second quantized theory in the sense that we describe states with an 
arbitrary nummber of strings, an achievement made earlier by String Field 
Theory by Kaku and Kikkawa and by Witten and others (for Bosonic string theory
\cite{Kaku3}). Our approach has some similarity to the work by 
C. Thorn \cite{Thorn} in as far as we also use a splitting of the string into 
small bits. 
We, however,  did it after the description of the 
single string has been resolved into right and left moving fields 
$X_R^{\mu}(\tau -\sigma)$ and $X_L^{\mu}(\tau +\sigma)$ (using the often used 
notation in string theory in so called conformal gauge) and most remarkably 
we achieve that we do not need any time development for what is the analogue of 
Charles Thorns small bits, we call ours as said somewhat different small parts 
for ``objects''. The ``objects '' stand, we could say, still\cite{self8}. 

It is this property of our string field theory which seems so interesting, that 
we indeed shall concentrate on precisely this no time development property 
in the present article.

We thus seek in the present article to go backwards in the sense that we begin 
by considering such objects, that do not develop, and then afterwards shall seek
by what we call ``phantasy'' to return to the string theory able to descrbe 
several strings. We believe it would be not so terribly hard to make the same 
``object'' description of the various superstrings, presumably by having 
``objects'' being  
 fermions with odd spin, but of course they should still 
not develop in time. Both because it would complicate the presentation and 
because we have not developped yet the string field theory of ours with such 
spin-objects yet - we did not even quite finish the bosonic version - 
we shall keep to the boson string theory in the present article.

The real motivation is to seek generalizations of the string theory. 
Indeed the main reason that superstring theory is so popularly speculated 
to be the ``theory of everything'' is that it avoids the ultraviolet divergence 
problems plaguing the point particle quantum field theories. This avoidance of 
ultraviolet divergences is connected with or a consequence of that the 
string scattering amplitudes - the Veneziano models - are strongly 
indeed exponentially cut off for large momentum transfer. In our ``object''
picture this cut off at large momentum transfer can be traced back to that 
the distribution of the momenta of the objects is cut off in a similar way.

The scattering in our picture of objects has the character of two 
composite particles($\sim$ strings) composed from the ``objects'' 
exchange some collections of objects with each other. Thus after the scattering 
the momenta of the final composite \underline{particles ($\sim$ strings)}
 can only 
deviate from those in the initial state by the momenta of the exchanged 
collections of objects. We shall namely remember the for this argument 
to be valid extremely important point that the single object never can 
change its momentum, because it does not develop at all, especially it 
does not scatter by itself. The whole seeming scattering of the composite
particles composed from objects is purely ``fake'' in the sense that they 
are a result of some objects being 
transfered from one compositum 
to another one.    

\vspace{0.5 mm}

(\ref{sec1} - 1 ){\bf Fake Scattering Concept}

We are so  fashinated by this idea of making a 
quantum field theory like theory, so that in a {\bf fundamental}
sense there is {\bf No Timedevelopment}, but when looking at 
it appropriately, then you can ``see'' it as e.g. string field theory
(a theory of second quantized strings).\\

This fake-scattering concept is implemented in the ``Novel string field
theory'' , which we have put forward long ago.


\vspace{0.5 mm}
(\ref{sec1} - 2 ){\bf Hamiltonian =0 gives no time-development}

So quantum mechanically the {\bf no time-development theory}
is just a Hilbert space of the states, and they never develop - there is 
basically no time needed -.\\
In the {\bf ``Novel string field theory''} of ours the states 
in this Hilbert space are described formally by a second quantized 
theory of particles that can occur in different numbers just like 
in usual second quantized theory. We call these particles 
``objects'' and they are crudely to be considered small pieces of 
strings like in the Charles Thorn's  string bit theory. But very importantly 
{\bf we } first split the string into bits, {\bf after we hav gone 
to the light cone varibles on the string}: $\tau-\sigma$ and 
$\tau + \sigma$.

\vspace{0.5 mm}
(\ref{sec1} - 3){\bf Introduction of Fake Degrees of Freedom}

In the philosophy that the true {\bf fundamental thory} has {\bf no 
time} (or say no time development) means that all development with time
has to be fake. That is to say it has to be in some degrees of freedom,
that do not really exist in nature, but which we the physicists introduce 
formally so as to make a theory more in agreement with our usual 
picture of how physics is.

Basically the idea is that we introduce some extra degrees of freedom 
that only are there in phantasy, so that we construc formally a system/a world 
with some extra variables or some extra information on its states. These extra 
inforations shall however only in some way be adjusted to help describing 
the original degrees of freedom, which we call the true degrees of freedom.
In the case we here hope to realize; the original or true degrees of 
freedom are the ones for the object. But by the addition of the extra 
degrees of freedom we have in mind giving an information about 
how the chains of the objects are glued together in possibly different ways.
These different ways we hope to describe by means of the extra phantsy 
degrees of freedom. It is mainly how the cyclically ordered chains 
of objects are, one can say the extra degrees of freedom should tell 
which objects belong to which cyclically orderd chain. Thereby the extra phatasy degrees of freedom also come to tell which objects belong to whch string.
Thus some strings exchanging objects can be a purephantasy happening.
This is what is called than scattering is a fake.  

Abstractly we replace each basis vector in a basis for the 
second quantized Hilbert space by a series of basis-vectors. 
So to each ``fundamental basis vector'' we have a lot of 
basis-vectors in the extended theory only deviating from each 
other by invented or fake degrees of freedom.

Then we allow the ``fake-development'' - the fake Hamiltonian -
to only move around the basis-vectors into each other which 
belong to the same fundamental basisvector.

(\ref{sec1} - 4) {\bf A String Field Theory Inspired Example}

To a good enough approximation the 
readers
can imagine that 
our ``objects'' (after some technical details of only using 
the ``even'' ones among them) are (scalar) {\bf particles} with 
position and momenta in a 25+1 dimensional world (or if we 
choose an infinite momentum frame in 24 transverse dimensions),
and that there in any single particle state for such a particle 
can be a number of particles $n=0,1,2,...$, just as in second quantization.

To avoid the problems with relativity, Dirac sea\cite{DS5} etc., 
we like to for pedagogical reasons effectively consider a 
non-relativistic theory, or almost equivalent an infinite momentum 
frame formulation.  

\vspace{0.5 mm}

(\ref{sec1} - 5) {\bf The Pedagogical Non-relativistic model with Zero Hamiltonian}

We consider a model with say non-relativistic bosons - so that 
they can occur in any number in any sigle particle state -.
To make the theory not develop in time we want to simplify to make the 
Hamiltonian zero
\begin{eqnarray}
H &=&0,
\end{eqnarray}
which in addition to having no interactions mean that we let 
the non-relativistic mass 
\begin{eqnarray}
m \rightarrow \infty,
\end{eqnarray}
so that even the kinetic term $\frac{\vec{p}^2}{2m}$ goes to zero.

We can choose a basis for the single particle states 
to be e.g. either the momentum eigenstates or the 
position eigenstates (a priori as we wish).

\vspace{0.5 mm}
(\ref{sec1} - 5 - 1){\bf Second Quantizing our $H=0$ Particle Model:}

As basis to use in single particle Hilbert space 
we shall here choose the position eigenstates because we like 
to investigate about a ``nearness'' concept (we want to say if two 
particles described by such basis vectors chosen are close or far 
apart.)

Then the corresponding basis in the second quantization state space
is enumerated by a function, that to every position $\vec{x}$ assigns 
a number $n(\vec{x})$ giving the number of 
particles with exactly the position $\vec{x}$.

In other words a second quantized basis-vector can be described by 
the number $n(\vec{x})$ of ``objects''(=particles) in each position $\vec{x}$:
\begin{eqnarray}
n : {\bf R}^{24} \rightarrow \{0,1,2,...\}
\end{eqnarray}
and we cannot require it continuos unless we take it to be only constant,
because a continuous function mapping the real number type of space
${\bf R}^{24}$ into a discrete 
space, the positive integers and 0,
can only be constant if it is continuous.

However, we shall at this stage  not describe  about continuity.

(\ref{sec1} - 5 - 2){\bf Second Quantized Basis}

 A basis - and this is the one we now have chosen to use -
in the second quantized state space consists of vectors like
\begin{eqnarray}
|n> &=& \prod_{\vec{x}}\frac{a^{\dagger}(\vec{x})^{n(\vec{x})}}{\sqrt{n(\vec{x})}}|n=0>
\end{eqnarray}
where $a^{\dagger}(\vec{x})$ is the creation operator for a particle 
at the position $\vec{x}$. The symbol ${\bf R}$ stands for 
the set of real numbers.

Remember
\begin{eqnarray}
n : {\bf R}^{24} \rightarrow \{0,1,2,...\}.
\end{eqnarray}

\vspace{0.5 mm}
(\ref{sec1} - 5 - 3) {\bf Introduction of the Fake Degree of Freedom ``The Successor 
Function'' $f$}

Our extremely simple $H=0$ theory just introduced has 
a priori nothing to do with strings (nor much other sensible 
physics for that matter), but now we want by just 
explaining to 
make it into a string field theory!

For each single one $|n>$ of our basis states in the second quantized space 
we want to introduce a ``sucessor function '' $f$, which is a 
permutation of the particles present in that state. 

In the state $|n>$ there are 
\begin{eqnarray}
N(n) &=& \sum_{\vec{x}}n(\vec{x}) 
\end{eqnarray}
particles present. Here we 
assumed that there were 
not infinitely many particles present. 

{\bf The ``Successor function'' $f$ is a Permutation of the Particles 
present in the state $|n>$.}

Assuming that there are only finitely many particles in a second quantized 
state vector $|n>$ we can think of these 
 $N(n)$ particles as true particles, and  you could 
define $N(n)!$ permutations $f$ of the $N(n)$ particles present.\\


\begin{figure}
\includegraphics{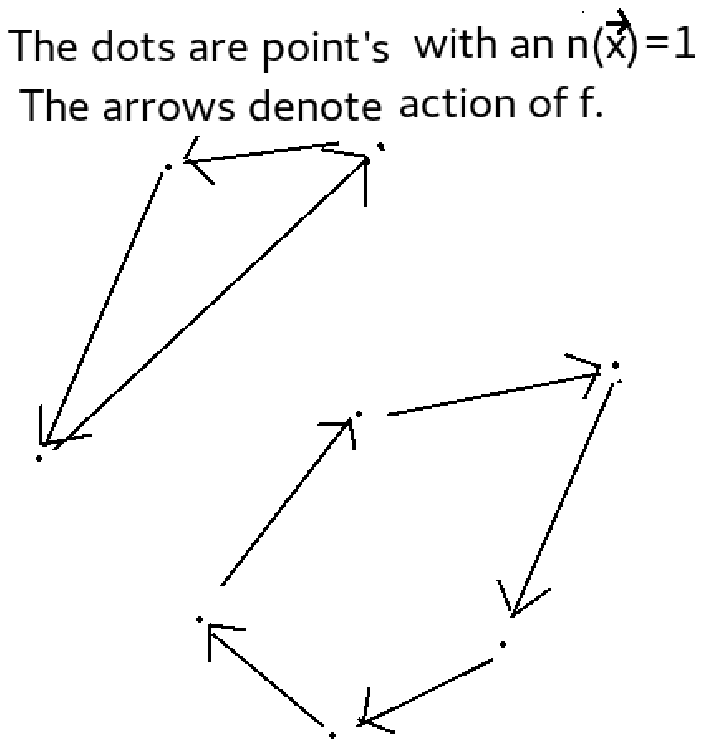}\\
\caption{The points denote the objects and the arrows denote the action 
of the permutation function $f$. Since a permutation can be resolved into 
cyclic permutations, the objects will on this drawing get into 
cyclically ordered chains.  }
\end{figure}

{\bf We Think of a Phantasy-space with $|n>$ replaced by $N(n)!$ new 
phatasy basis vectors 
representing the same true physics.}

So the new basis-vectors in the second quantized space should be 
denoted
\begin{eqnarray}
|n, f> &=& \left ( |n> , f \right ) \hbox{ where } f\in P_{N(n)}
\end{eqnarray}
where again 
\begin{eqnarray}
N(n) &=& \sum_{\vec{x}}n(\vec{x}) 
\end{eqnarray}
is the number of particles in the state $|n>$.

{\bf Working with Phantasy space Makes Life Easier}

Of course it is $f$ which is the phantasy degree of freedom.
It was just introduced by us.

"Thus we can decide in the following rule:

We throw away all  the choices of permutations f unless it fullfils 
the following rule
(we do so for some reason to be explained later):
The position $x_{first}$ of a particle 
$first$
being mapped by $f$  to $f(fisrt)$ must have 
a position $x_{f(first)}$ close to $x_{first}$. 
That is to say,
we require only to include in our phantasy space
such combinations that $ f$should satisfy f(first) is close to first,
i.e. $| x_{first} - x_{f_f(first)}|$  should small.

If $f$ does not obey this restriction,we simply take it out
and let there be fewer state vectors in the phantasy Hilbert space."


{\bf We can phantasize that $f$ describes successors in long 
almost connected chains}

We can choose the $f$ permutations, we allow, to be such that they describe 
connected closed loop chains of the particles in the 
state, so well it 
is possible.\\
 

From our purpose of making theory to be part of a speculated 
theory for everything we could be allowed to postulate something 
- if beautiful enough - also {\bf about the state of the universe},
such as that the most liklely type of state is one in which the 
particles sit in long circular chains with rather small
distance between the neighbors and even further assumtions 
involving the momenta.\\



\begin{figure}
\includegraphics{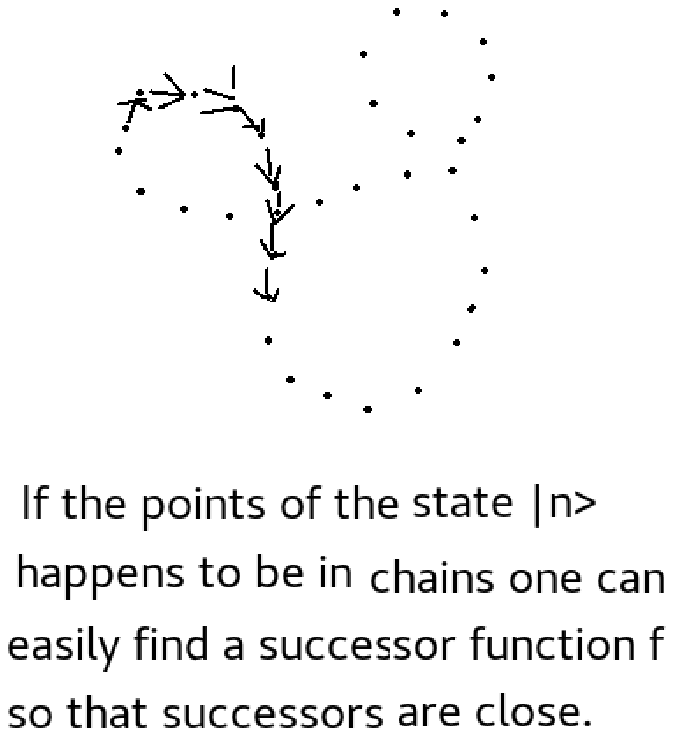}
\caption{We think of the objects sitting (due to some assumed 
principle about the liklely configuration of the objects) in cyclically 
ordered chains, that can be described by giving the permutation function 
$f$ as here illustrated by the arrows. The reader can imagine contuing 
marking the arrows to tell how $f$ acts.}
\end{figure}

\vspace{0.5 mm}
(\ref{sec1} - 6) {\bf About Fundamental Physics We can further only make 
assumptions about the 
Initial and /or Final states}

After we settled on no time-development ( $\sim$ Hamiltonian being zero)
we can {\bf not} as physicists looking for the right theory of nature anymore 
{\bf speculate about the Hamiltonian}, because that we already took to zero
(as operator).

But we may {\bf want to} have a bit of chanse to {\bf assume} a little bit 
to adjust to fit our hoped for model to experimental information etc.

Then we have the chance of speculating about the {\bf initial state}
(which is also the final state though, when no development).

{\bf Assumptions about Initial and final state}

For our purpose with the string theory towards which we are 
driving in mind we like to make assumptions about initial condition 
like this:

\begin{itemize}
\item{(a)} An approximate constraint on the relative state of a couple 
of particles $A$ and $f(A)$, namely
\begin{eqnarray}
k( \vec{x}_{f(A)}- \vec{x}_{A}) &\approx & \vec{p}_{f(A)}+ \vec{p}_{A}, 
\end{eqnarray} 
where $k$ is a constant, actually related (as to be seen) 
to the Regge slope $\alpha'$ so important in string theory.

\item{(b)} The particles shall approximately form cyclic chains.

\item{(c)} And they shall even especially locally along the chains 
have a certain wave function like they would have 
in string theory if they were identified with the ``objects''
of ours (which we have not yet described in detail.) 
\end{itemize}

(\ref{sec1} - 7) {\bf Assumptions about (Initial) State Formulated by Density 
Matrix $\rho$}

Whatever assumption about a quantum system one might want to make
it can in principle be written by means of a {\bf density matrix } $\rho$.

$\rho$ is a positve operator on the Hilbert space of state vectors for the 
system normalized to $Tr(\rho) =1$.

We have one $\rho_{fundamental}$ for the ``fundamental degrees of freedom, and we 
can partly choose one $\rho_{full}$ for the combined system of the fundamantal 
and the phantasy degrees of fredom system. Then you can act 
\begin{eqnarray}
&& \rho_{fundamental} |n>\nonumber \\
\hbox{or}&&\nonumber \\
&&  \rho_{full}(|n>,f)\nonumber 
\end{eqnarray} 

{\bf Density Matrix Relation}

We shall naturally require for consistency
\begin{eqnarray}
<p|\rho_{fundamental}|n> &=& \sum_f (<p|,f)\rho_{full}(|n>,f)\\
\hbox{or formulated differently: }&&\nonumber \\
\rho_{fundamental} &=& Tr_{w.r.t. \; phantasy}\rho_{full} 
\end{eqnarray} 
So far we talk about timeless density matrices.

But could we make a purely phantasy time development of only the 
phantsy or f-degrees of freedom without distrubing the fundamental 
( $|n>$ , $|p>$,... ) degrees of freedom ?


{\bf Stringy Initial State Assumptions, and Phantasy Notation give 
String Field Theory}
The point is we put a fairly large  amount of {\bf string theory into assumptions 
about the initial state}, partly because we cannot do it  in the 
proper Hamiltonian.

\underline{The assumption},

 ``An approximate constraint on the relative state of a couple 
of particles $A$ and $f(A)$, namely
\begin{eqnarray}
k( \vec{x}_{f(A)}- \vec{x}_{A}) &\approx & \vec{p}_{f(A)}+ \vec{p}_{A}, 
\end{eqnarray} 
where $k$ is a constant, actually related (as to be seen) 
to the Regge slope $\alpha'$ so important in string theory.''

This assumption would if the particles did not have infinite masses mean that 
the cyclic chain would move along itself. 


(\ref{sec1} -8) {\bf Yet a complication in relating the trivial static theory to string theory}

The cyclic chains of particles are {\bf not} simply the 
strings when we identify with string theory - as it would be in 
Charles Thorns theory -, No. 

We have to choose a starting point and go along the 
cyclical chain from that with two marks in opposite directions 
along the chain, and then construct for each step an average 
of the two ``poeple'' that started at the start. It is the series 
of average under this trip of the two ``people'' that makes up the string.

In this way 
we get an open string from making this two ``poeple''
walk on a cyclically ordered chain. 

\begin{figure}
\includegraphics{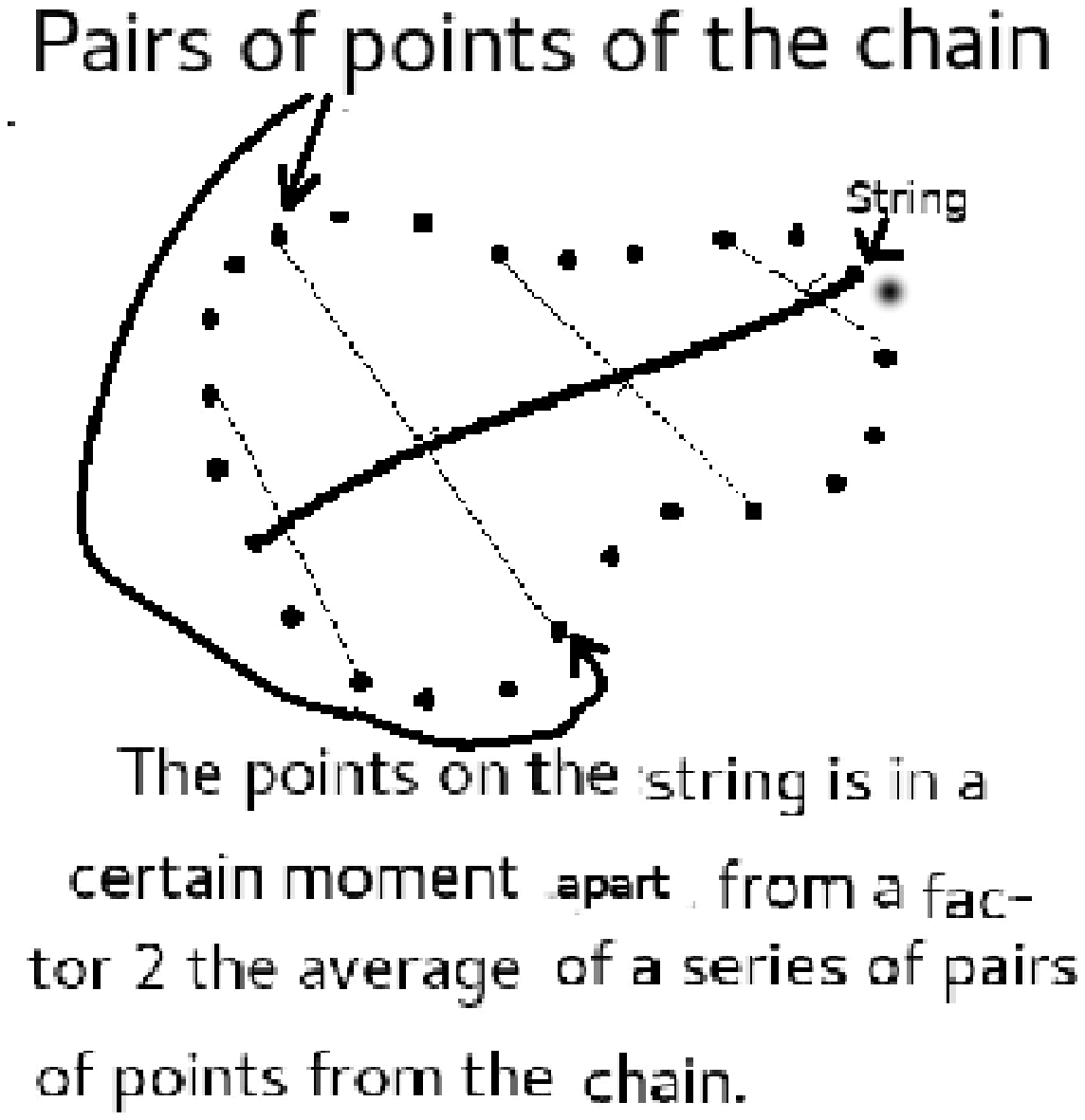}
\caption{How to construct an open string in terms of objects:
You need only one cyclically ordered chain for giving an open string
with apart from a factor 2 having the points of the string being the averages 
of the two object points in the pair. Of course one can pair the objects in 
different ways even keeping to the continuous type of way illustrated, and 
that then give the string at a different moment of time.
To ensure that the reader identifies the right small spots with the objects 
one may count that there are 26 objects on this drawing. Corresponding to that 
there must be 14 points on the open string. The long curved arrows just point 
to two objects forming a pair, but here are 14 ``pairs'',corresponding to 
14 points on the string, the objects at 
the ends of the string being paired with themselves.  
}
\end{figure}

\
\begin{figure}
\includegraphics{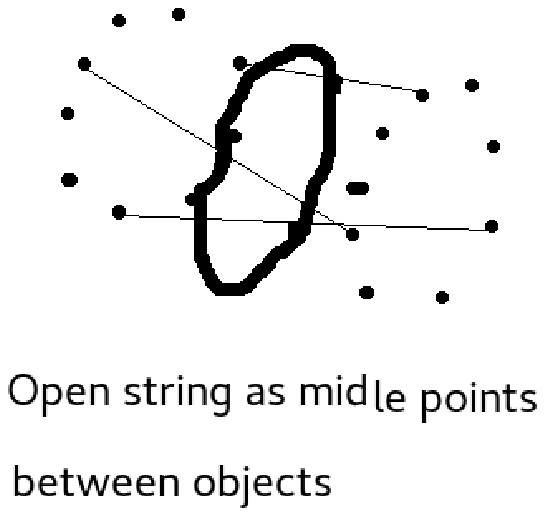}
\caption{How a closed string is constructed from objects: You need for that
{\bf two cyclically ordered chains} of objects (as ordered by $f$, but $f$ is
not drawn on this figure) and except for normalization by a factor 2 you 
construct the averages for pairs of objects with one object in the one 
cyclically ordered chain and the other member of the pair in the other 
cyclically ordered chain}
\end{figure}


{\bf Main Point: Brought although a bit complicated a correspondence to 
String Theory}

To a set of strings in a known state - e.g. the ground state of their 
oscillations - one can calculate the state of the correponding 
particles (which we usually call ``objects'') sitting - ordered by the 
faked $f$ description - in a cyclic chain for each \underline{open string} 
(we postpone the \underline{closed strings} for the moment). 

I.e. {\bf We can pretend to see string theory in our game with infinitely 
heavy particles.}

Most remarkably: When we calculated the overlap between two different 
sets of strings represented as second quantized states of objects, we got 
- apart 
from a wrong sign (a missing i) - the form of the {\bf Veneziano model}.

As the typical example we considered an initial state with two 
cyclically ordered chains of objects representing two open string 
in the states of the ground state of bosonic strings. Then as final 
state we took a similar state of objects correponding to two ground state 
open string. We allowed, however, these open strings to have arbitrary 
momenta. Then the overlap indeed run out to become the four point Veneziano
model for the two incoming and two outgoing particles identified with 
the strings. We did though get two ``small'' problems: 1) We had expected 
to get a Veneziano scattering amplitude being a sum of three Beta-function 
terms, but we got only one term.  2) If we should expect the overlap 
we calculated to be identified with the S-matrix element - as the S-matrix in 
theory without time development would be expected to be 1 - we should have 
gotten the Veneziano model with an extra factor $i=\sqrt{-1}$ because there is 
in the expression for the S-matrix in terms of the amplitude which is real in 
lowest order Veneziano model with an extra $i$.

But that was what we got at first. 

{\bf Correspondence with Veneziano Model rather short via thinking on surfaces
of string development}

\begin{figure}
\includegraphics{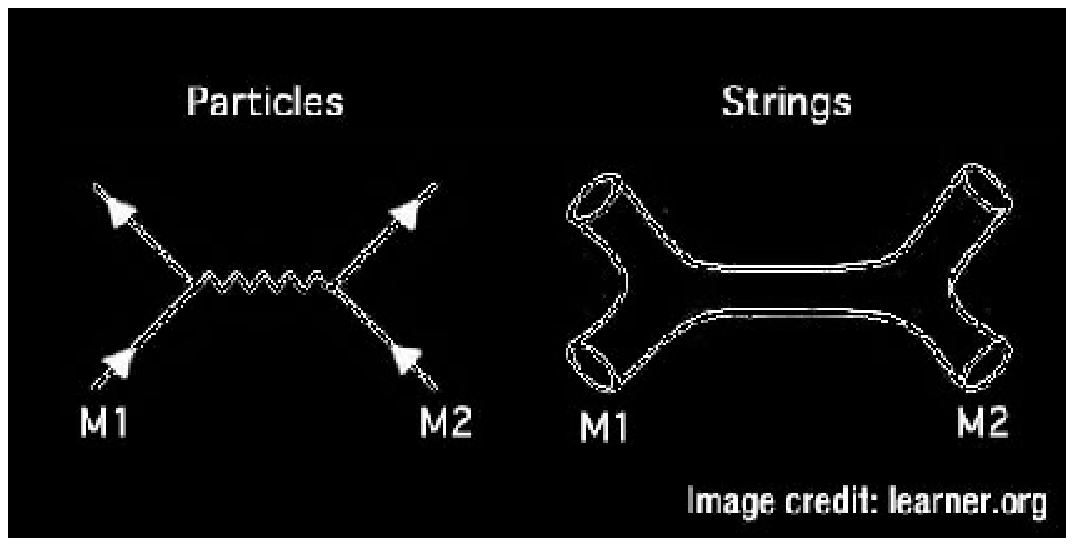}
\caption{}
\end{figure}


(\ref{sec1} - 9) {\bf Important step in Showing Veneziano Model from Our 
Novel String field theory}

You think of external ground state strings. They can be produced 
as in general ground states - by a long imaginary time development with 
the appropriate Hamiltonian. This development is then written as in complex 
time development of the string, very reminiscent of what it always used 
in string theory to compute say Veneziano model.

Very crudely we just give a motivation for this kind of functional integral 
description of the strings.

Really we do it with a doubled string; i.e. we have a closed  
string diagram describe the open string. So there are some complications 
but we did mannage to one of the three terms.

{\bf Changing Phantasy Degrees of Fredom can Change Number of Cyclic Chains 
and thus of (Open) Strings}

For different ``successor functions''  $f_1$ and $f_2$ you can find 
different numbers of cyclic chains even for completely the same 
configuration of the infinitely heavy particles (=``objects'') 
and thus in fundamental physics-wise the same situation $|n>$. 

{\bf Take a fundamental physics situation $|n>$ like this:}
\begin{figure}
\includegraphics{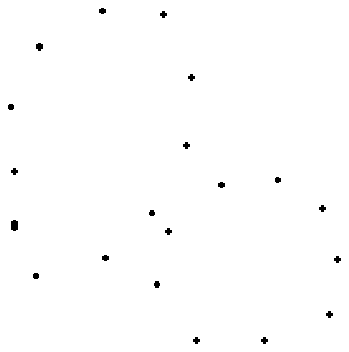}
\caption{}
\end{figure}

{\bf In choice $f_1$ of the phantasy you have one open string}

\begin{figure}
\includegraphics{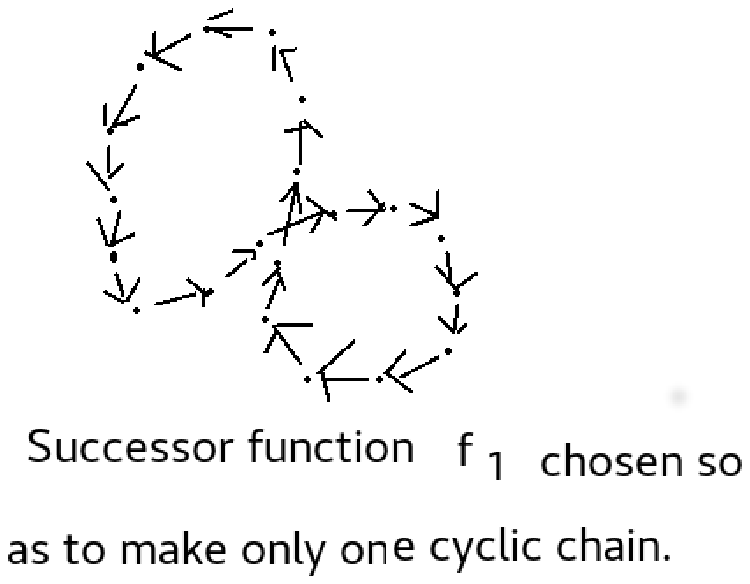}
\caption{}
\end{figure}


{\bf In another choice $f_2$ of the Phantasy gives Two chains, thus Two open 
strings}

\begin{figure} 
\includegraphics{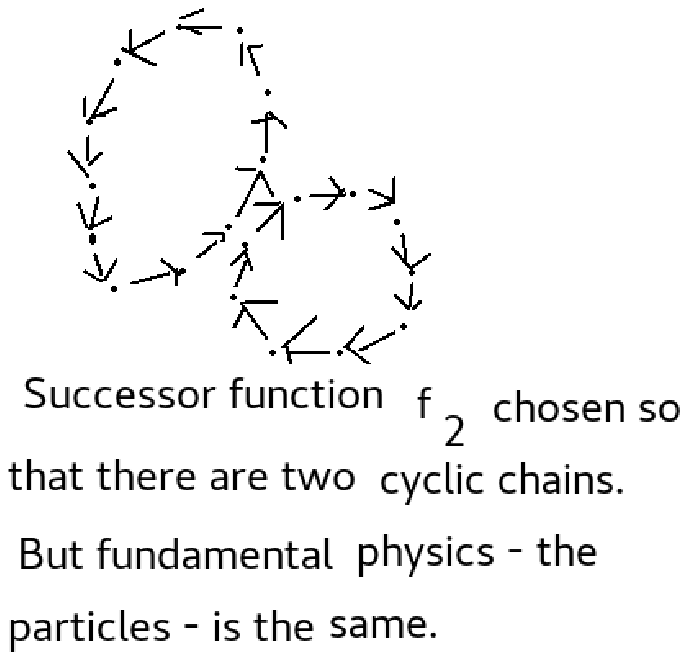}
\caption{  On this figure the reader sees the same points illustrating the 
``objects'' as on foregoing figure, and we hope the reader can see that 
one can have two seperate cyclically closed chains that though come very close 
to each othet at some point. It is such cases that the successesor function 
$f$ can be change a few places and still be in agreement with the 
approxmate requirement that the successor function maps one object to the 
next in a chain to which it belongs.}
\end{figure}


{\bf Unification of strings can be change of $f$, thus phantasy}, because 
changing actually actually only a couple of values of the successor function 
$f$ can cause that e.g. a previously closed cyclically ordered chain of 
objects get split into two such chains. Since this would correspond to one 
open string being split into two, we see that splitting can be easily described 
by changing $f$. Oppositely of course the oppostite change in $f$ would mean 
a unification of two to become one openstring.

\begin{figure} 
\includegraphics{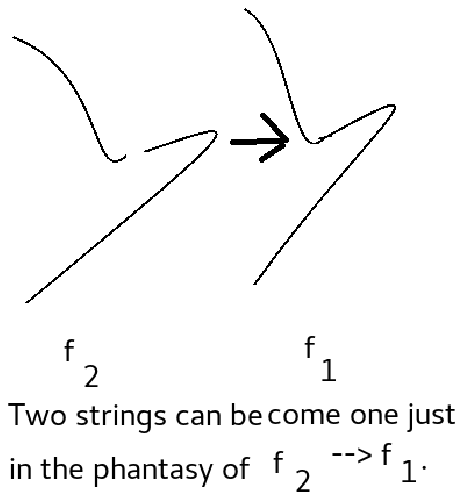}
\caption{}
\end{figure}


\section{Hamiltonian}

{\bf How to make a purely Phantasy Hamiltonian 
}

The excercise we want to do now is to see what Hamiltonian 
is allowed working on the extended Hilbert space containing also the 
phantasy degrees of freedom, so that the basis states are 
\begin{eqnarray}
\hbox{(\underline{Extended}) Basis states } &\hbox{of the form }& (|n>, f)\\
\hbox{while}&&\nonumber \\
\hbox{\underline{Fundamental} basis states }&\hbox{of the form }& |n>
\end{eqnarray} 

Thinking of matrices the extended operator (matrix) consists 
of a lot of blocks, one block for each matrix element in the 
(original) fundamental Hamiltonian (which is actually zero). 




{\bf Attempting to find a phantasy Hamiltonian only moving the Phantsy 
Degrees of Freedom}

For any operator depending only on the physical degrees of freedom 
$O$ we want the ``only phantasy'' Hamiltonoian $H_{phantasy}$ candidate to 
commute with it:

\begin{eqnarray}
[ O, H_{phantasy} ] &=&0.\label{eq17}
\end{eqnarray}

This condition is, however, too strong, since it would not allow the 
hamiltonian to depend at all on the ``fundamental'' degrees of freedom, 
because if so, the conjugate variable to the one it depended on would 
be made to vary (and that we wanted to exclude).


 We must be satisfied with only having this requirement {\bf approximately.} 


{\bf Not so good Argument that we can have a wanted $H_{phantasy}$ approximately.}

We want the development in the $f$ or phantasy degrees of freedom only 
to depend on that some cyclic chains come very close / touch; and that is 
dependent on only very few particles/``objects'', so at least it does 
only involve at first few among an extremely big number of ``objects'' 
in the interesting situation.

\subsection{An Attempt to an $H_{phatasy}$}
\label{subsec21}
If one thinks of using 
position eigen-state say, one could 
enumerate the particles by the position $\vec{x}$ 
or we could say that the number $I$ were a function of the 
position $\vec{x}$ defined for the values of the latter
a particle, i.e. defined whenever $n(\vec{x}) =1$ or bigger. 
In the case when there are positions with more than one particle the particle number $I$ would in addition 
have to depend on a small $i$ number eneumerating the particles at one 
special position $\vec{x}$ say. So we would in this more general case 
write 
\begin{eqnarray}
I &=& I(\vec{x}, i)= \hbox{The number $I$ of the $i$'th particle 
at the position $\vec{x}$.} 
\end{eqnarray}   
where the $I$ is then a name or a number assigned to the $i$th object 
at the position $\vec{x})$.
In order that at least Hamiltonian $H_{phantasy}$ should commute with 
those operators $O$ that could be constructed out of the 
position operators of the ``objects'' alone, we would have to make 
the $H_{phatasy}$ operator at least not change the second quantized state in the 
position basis formulation, when acting on it. As soon as we would make an 
image of $f$ say $f(I)$ to be change to be an object sitting at a different 
position than before the action, this could give rise to the change 
by the action of some only on the positions depending operator, and thus 
this would not be allowed.

This consideration would leave us with only the possibility that the 
action with $H_{phantasy}$ to change the image $f(I)$ from what it starts being 
to the number for an object with the {\bf same position.}. That is to say
that if say 
\begin{eqnarray}
f(J)&=& I(\vec{x},i)\hbox{before the action with $H_{phantasy}$,}\\
\hbox{ then }&&\hbox{ after the action:}\nonumber \\
f(J) &=& I(\vec{x}, k) \hbox{ (where $\vec{x}$ is the same, but $k$ 
can be different from $i$.)} 
\end{eqnarray}

This means that under the operation of the phantasy Hamiltonian 
we can as far as these $\vec{x}$-representation considerations go
change the $f$ into another $f$ let us say $f'$ obtained by 
multiplying it - in the permutation composition way - from the 
right 
by a permutation of a subset of objects sitting on the 
same position. If $P_{\vec{x}}$ denotes the subgroup of the permutations
of the objects with position $\vec{x}$, we can say it would be allowed 
that the 
sucessor function $f$ changes into $f'$ with 
$f'= f\circ p$ for some $p\in P_{\vec{x}}$ for some position $\vec{x}$.

That is to say that worrying only about the (fundamental)
position dependent operators w.r.t. whether they commute with  
the phantasy hamiltonian we can allow matrix elements like :
\begin{eqnarray}
<n,  f\circ p|H_{phantasy}|n, f> &=& g \hbox{ (some nonzero value)} 
\end{eqnarray}
where $p$ is a permutation of objects with the same position.
(The value $g$, which we must here introduce, will turn out to be 
proportional to the coupling constant, also usually called $g$ 
in the formulation of veneziano models.). 

This proposal is, however, although it looks at first o.k. {\bf not good}:
The point is that if we concentrate on some successor function $f$ having 
resulted by mutiplication with such a permutation $p$ then you will in that 
state find that there is an infite uncertaity in the relative momentum for 
the two objects that had the same $\vec{x}$ position. It would act much like 
they had scattered with a pointlike interaction. This would mean indeed that 
the $H_{phantasy}$ had changed the state of the fundamental degrees of freedom 
and we wanted to avoid that. Because if our playing or phantasy Hamiltonian 
truly change the state of the tue funndamental degrees of freedom it is 
not truly only phantasy.

The occurence of the need 
for selecting a permutation of the objects sitting on the position $\vec{x}$ 
say a priori is a little freedom to be specified, but in what we 
think should be the most important situation:
\begin{itemize}
\item{(\ref{subsec21}.a)} That there are in most positions $\vec{x}$ no 
objects at all,
i.e. most $n(\vec{x})=0$.
\item{(\ref{subsec21}.b)} and the dominant part of the rest of the positions 
have just 
one object, i.e. next to $n=0$ it is the value $n=1$ that is most common.
\item{(\ref{subsec21}.c)} Continuing this way with falling numbers of 
positions the higher $n$,
the first and dominant value of $n$ for which a non-trivial permutations 
of the objects at the position is $n=2$ for the position in question.
And in this case there is {\em only one nontrivial} (i.e. not 
unity) {\em permutation of the objects at the position}. So in this 
most copious non-trivial case the permutation $p$ is not ambiguous.
\item{(\ref{subsec21}.d)}  Finally we expect higher $n$-values than 2 to be 
extremely seldom, and we 
may ignore approximately this possibility.
\end{itemize}  

Of course it is so to speak the choice of the density matrix $\rho$ which 
shall give the a priori probability for how to find the system of objects 
that should be made so that we have this probability for $n$ taking 
a given value to fall rapidly with the size of this value. 
It is actually very natural with such a property in the limit of the 
$\vec{x}$-space going to a continuum.

The reader may check that we also without causing any problem 
for the conservation under the phantasy Hamiltonian development 
of the operators depending on the positions for the objects by also allowing
 \begin{eqnarray}
<n,  f|H_{phantasy}|n, p \circ f> &=& g^* \hbox{ (the complex conjugate of $g$)}, 
\end{eqnarray}
so that we can achieve that the phantasy Hamiltonian is a Hermitean 
one in the phantasy Hilbert space constructed as tensor product of the 
space with the $f$'s as basis vector marks and the fundamental Hilbert space.
So we can claim we arrange:
\begin{eqnarray}
H_{phantasy}&=&H_{phantasy}^{\dagger}.
\end{eqnarray}

\subsection{The problem of keeping fundamental degrees of 
freedom fixed}
But this proposed $H_{phantasy}$ will not commute with the relative momentum 
of the typically two  objects being permuted by $p$, because the 
expression we proposed depends on the relative position and thus will not
commute with the conjugate momentum. 

Actually as already said it is impossible to solve this problem except at best 
approximately somehow. If we truly arranged that the phantasy Hamiltonian 
should commute with all operators from the fundamental degrees of freedom, we would be forced to have a phantasy Hamiltonian only depending on the 
pure phantasy degrees of freedom, and that would not be so fun.

But we anyhow want to speculate that such a phantasy Hamiltonian can 
act approximately without disturbing the fundamental degrees of 
freedom significantly. E.g. one could 
speculate that as described in the now 
following subsection, it would approximately commute enough 
under assumption of the density matrix distribution. 

\subsection{A rather bad example for idea of concrete $H_{phantasy}$}

Since we actually have just seen that a fully satisfactory phantasy Hamiltonian 
is impossible (see the argument below formula (\ref{eq17})) , just propose 
one that has difficulties in the sense 
of not commuting fully with the fundamental degrees of freedom -
meaning operators acting only on the original basis $|n>$ space -
would still be of interest. To give the possibility to work 
on with the idea of constructing a phantasy Hamiltonian that functions 
approximately let us indeed build the proposal from an operator 
$N(I,J)$ supposed to act on the space of fundamental states and 
being effectively zero in all cases when the objects $I$ and $J$ 
are not close to each other and only significant when these two 
objects are close to each other. You may take it that it is 
so to speak a ``nearness operator'', and that  is why we called in by the 
first letter $N$ in the word ``nearness''. Such an operator $N(I,J)$
is to be considered an operator of the same kind as an interaction 
between the two objects $I$, and $J$, and thus could be written as a 
convolutions by some function possibly involving smeared delta functions 
The operator $N(I,J)$ of course only act on the wave 
functions for just those two objects $I$ and $J$, so it could be written 
acting on the space of all the objects represneted by wave function like 
$\psi(x_1,...,x_N)$ as 
\begin{eqnarray}
&&N(I,J) \psi(\vec{x}_1,...,\vec{x}_N)= \nonumber\\ 
&=& \int\int d\vec{x_I'}d\vec{x_J'}
K(\vec{x}_I,\vec{x}_J;\vec{x}_I',\vec{x}_J')
\psi(\vec{x}_1,..,\vec{x}_{I-1},\vec{x}_I',\vec{x}_{I+1},...,\vec{x}_{J-1}, 
\vec{x}_J', \vec{x}_{J+1},...,\vec{x}_N)\nonumber
\end{eqnarray}   

Of course in order that $N(I,J)$ be a nearness operator the 
to be chosen function of four spatial vector $K(\vec{x}_I, \vec{x}_J;
\vec{x}_I',\vec{x}_J)$ should  vanish for 
any of the four arguments being far away from the other ones.

We should also think of it as being in spite of its locallity 
rather smooth so that it does not change the momenta of the objects $i$ 
and $J$ too much. Actually the reader should understand that we are hoping 
for - the impossible - that we have an operator just testing if the 
two objects numbered $I$ and $J$ are near each other, but preferably 
without disturbing them. But we know from the discussions 
of Niels Bohr etc. that in quantum mechanics you cannot measure 
without disturbing. 

Anyway let us go on for pedagogical reasons as if we had arranged
an operator $N(I,J)$ that could just observe without disturbing.
This would be an only classical intuition that could have that.

If we anyway fall back on classical intuition, we could as well 
really take it as if the operator also asked for {\bf nearness in 
momentum space}, i.e. it should be arranged to only be significant in size 
for the two objects having approximately the same momenta also.

Supposing we now had a for practical purposes such nice 
operator checking if two objects are in the approximately same 
point in the phase space $N(I,J)$.

The idea then is that we shall by means of it construct a term 
in the phantasy Hamiltonian that if $N(I,J)$ is non-zero will 
permute the actions of the successor function $f$ on the two objects 
involved. That is to say that with a weght $N(I,J)$ the successor function
$f$ shall be changed so that the images of $I$ and $J$ are no longer 
as at first $f(I)$ and $f(J)$ respectively, but oppositely 
$f(J)$ and $f(I)$.

This means that we define a term to be put into the phantasy 
Hamiltonian
\begin{eqnarray}
H_{IJ} &=& N(I,J) P_{f \rightarrow f\circ p_{IJ}},
\end{eqnarray}
where $  P_{f \rightarrow f\circ p_{IJ}}$ is an operator only 
acting on the phantasy-degress of freedom , i.e. on the 
f-part, by permuting the two object(numbers) $I$ and $J$
before the action of $f$. Here the permutation $p_{IJ}$ 
means the permutation permuting the two objects $i$ and $J$.

The full proposal for the phantasy Hamiltonian should then be the sum 
over all pairs of different objects $(I,J)$, with $I ne J$.

That is to say we propose the phantasy Hamiltonian to be of the form:

\begin{eqnarray}
H_{phantasy}&=& \sum_{(I,J)\; with \; I\ne J}H_{IJ}\\
&=& \sum_{(I,J) \; with \; I\ne J}N(I,J)  P_{f \rightarrow f\circ p_{IJ}}.
\end{eqnarray}   
This phantasy Hamiltonian is made so as give some topology 
change - change in the way the objects are thought to hang 
together in chains (the cyclically ordered chains) as the 
phantasy-time goes on, but only provided the chains almost 
coincide where the change takes place. This will corrspond also 
when translated into strings shifting the topology of how they 
hang together to only glue the strings in a new way at 
places where they touch. This is what you expect for 
physical strings also: they only interfere when they touch.

As already stressed the $H_{phantasy}$ here is at best approximately 
o.k.. We can argue that it is not so bad again by remarking that 
if one thinks on strings with infinitely many objects in them and that 
we can arrange the interaction between the strings to be sufficiently 
weak - by putting the coupling $g$ above absorbed in $N(I,J)$ 
sufficiently small - so that one only has about one interaction at a 
time, meaning that only one out or two of infinitely many objects
get disturbed by the operators $N(I,J)$.
 
\subsection{Mathematically Formulated Approximate 
$H_{phantasy}$ Restriction}

One idea to make a concrete statement of the sought for purely 
phantasy hamiltonian, that should preferably only move the 
phantasy degrees of freedom $f$ but not the fundamental degrees 
of freedom $n$, would be to replace the hoped for 
$[O,H_{phantasy}]=0$ requirement by the milder
\begin{eqnarray}
Tr(\rho [O,H_{phantasy}]) &=& 0,\label{densitycommutator}
\end{eqnarray}
for all genuinely fundamental operators $O$ and the assumed 
density matrix $\rho$ expressing our assumtion about the state of the 
system of ``objects''. We should presumably most 
wisely only take this relation in the limit of infinitely many 
objects and then we can hope as just mentioned that a single object 
being a bit pushed would not count very much, if it stands inside 
the quantum fluctuations.

It is easy to see by a bit of trivial algebra that if we 
choose $H_{phantasy}$ to commute with the density matrix 
$\rho$ we get fullfilled (\ref{densitycommutator}).

This means that we should look for arranging that our 
$N(I,J)$'s in the phantasy Hamiltonian commute with the density 
(matrix) operator $\rho$.

\subsection{Unitarity}

Once we have settled on a formalism with a constructed phantasy 
Hamiltonian, we can of course construct corresponding time development 
operators, say the time-development operator from time $t_1$ to 
time $t_2$ would be 
\begin{eqnarray}
U(t_2,t_1) &=& \exp(-i H_{phantasy}),
\end{eqnarray}
(of course a phantasy development). This time-development 
- which is also an approximate S-matrix - would of course 
be a unitary operator acting in the space extnded with the phantasy 
degree of freedoms. Thinking of the development with lowest 
order perturbation in the parameter $g$ leading to the Veneziano model 
as we have previously argued, it is essentially obvious that the 
higher orders will give unitarity corrections to this Veneziano model.
So the scheme with the phantasy Hamiltonian should automatically lead to 
include these Veneziano model unitarity corrections.

( Let us though at this point remind of the problem we had in deriving 
the Veneziano model: when we did it in the infinite momentum frame
- which is very close to the non-relativistic game used in this article - 
we did not get but one of the three terms we ought to have got. Of course then 
we shall also miss some of the unitarity correction terms if we just use the 
here a bit simplified form.)

\section{On the Concept of Time.}
As a little parantesis at this point let us point out that our 
picture with the stressing of no time development, really means that 
in our object-description there is at first {\em no time}. One can say 
that the time first comes in when we introduce the phantasy degrees of freedom,
and the phantasy Hamiltonian. In this sense the concept of time comes 
into our scheme as a ``phantasy'' a fake. The fundamental world 
has no time. Only by looking at situations in which 
the various pieces of cyclic chains are screwed together in deifferent 
combinations as existing at different moments of time a 
time-concept pops up. That is to say that if one wants to make some 
ontological model for how a concept of time comes into physics, then 
we here have the roots for some idea about that:

The time concept could be a phantasy degree of freedom which for some 
reason could be a reasonable way of describing an a priori timeless 
physics.

Interestingly enough this attitude of time being a phantasy or fake 
concept is not actually quite new in as far as we can 
claim that it is already present in general relativity:

In general relativity all the coordinates and not only (but also)
the time coordinate $t=x^0$ are arbitrary and phantasy or 
fake, in the sense even that the physicist that chooses the coordinates,
can decide what these coordinate shall be.

Crudely imposing quantum mechanics and reaching the Wheeler-DeWitt 
equation one has by this Wheeler-DeWitt equation a restriction 
on the state, which seemingly tell that the state of the gravity 
theory is the same at all times. The most close to a Hamiltonian in 
the gravity theory is namely an integral over the Wheeler-DeWitt equation 
quantity. This then means that one has got a constraint that the Hamiltonian 
shall be zero as a constraint. So taking this at face-value one has 
in gravity a  very similar situation as to the one in our scheme:
There is no time development, except in some gauge-chosen or fake way. 

\section{Motivations}

{\bf Purpose of this Faked Scattering String Theory Formulation}

Hope you got the idea of {\bf considering a completely trivial 
$H=0$ quantum field theory} and built up a story of 
e.g. strings {\bf just by defining some extra ``phantasy degrees of 
freedom''}.

What is the purpose ?:

\begin{itemize}
\item{(4-a)}  It is a method to make a second quantized string theory 
(competing 
with works by Kaku and Kikkawa and by Witten, ...\cite{Kaku3}). You can describe states 
with several strings.

\item{(4-b)}  You may use the idea to look for further models sharing the 
great property of string theory of {\bf not having the usual divergencies}.
Likely this is the only hope for making theories, that make sense, in high 
dimensions.

\end{itemize}

{\bf Problem of Ultraviolet Divergences Worse the Higher Dimension of 
Space-time}

Each momentum-formulated loop intergal in a Feynman diagram bring a 
$ \int ...d^dq $ integration and unless there are very many propagators 
in the loop we cannot avoid divergence for large loop momenta $q$.

\vspace{0.2 cm}

The higher dimension the more different loop integrals lead to 
divergencies. 

\vspace{0.2 cm}

To absorb the divergencies into bare coupling constants you need in 
high dimensions so many that the theory ends up with infinitely many 
parameters, and is in principle useless.


{\bf Direction of Hope for High Dimensional Theories: Formfactors}

One needs some factor that can make converge the loops in the high dimensional 
theory, otherwise you have ultraviolet divergencies and in high dimensions
it gets too many different divergencies.

\vspace{0.2 cm}
Best hope:

{\bf some exponentially falling off factor}
\begin{eqnarray}
\hbox{Factor extra in loop} &\propto& \exp(- k*q_E^2)
\end{eqnarray} 
much like what one gets from formfactors when one has effective theories 
for hadrons. \\

\vspace{0.2 cm}
Suggestion:

Replace the particles in the high dimensional theory by 
{\bf composite (bound) states}, like the hadrons are composite in 
QCD. 


{\bf Just Bound States Not Good Enough: Partons\cite{Bjorken6} }

If as we now believe hadrons are bound states\cite{BS} but of quarks and 
gluons called in this connection partons the effective vertices 
will NOT go down exponentially for very big momentum transfers 
but will be dominated by the coupling to a single parton and behave at 
the end more like in the theory of just particles. Thus it will only 
help a part of the way, but finally at high momenta the divergencies 
reappear. 

{\bf Only if there are infinitely many consituents(=partons\cite{Bjorken6}) in the
bound states and they have Bjorken variable $x=0$, you can postpone 
parton dominance from popping up, and thus only then we can use the 
replacement of the original particles in high dimensional theory 
by bound states.}


{\bf Hadrons Scatter Crudely by Exchange of Bunches of Constituents}

Hadron scattering at energies below where partons 
collisions become important was described by exchange of other hadrons,
pions, vector bosons like $\omega$, again hadrons which again consists 
of many partons. So it was mainly exchange of lots of partons
between one hadron and another one, while the single partons hardly were seen.\\

\vspace{0.2 cm}

{\bf Moderate energy Hadron scattering in terms of partons
is much like the fake-scattering  of just exchanging bunches from 
one bound state to another one.}

(we here ignored the relativity and effects of vacuum)

\subsection{Bound States Not Perfectly True for Our Fake Model}

Let as a side remark call attention to that our model of the string 
in string theory as ``composed'' of constituents which we called 
``objects'' should presumably not really be called composite in as 
far as when we stress that there is no interaction it is not truly 
a bound state. You could of course think that one could the limit of 
letting the interaction be weaker and weaker and thus at the end have the 
non-interacting consituents. One could think of some weakly bound states 
such as some molecules or atoms and then consider a process in which 
- for some reason a very fast  - exchange of say an electron or 
some other combination of electrons and nuclei, e.g. some whole atom 
takes place between a couple of different molecules. If this exchnge goes 
fast compared to the internal 
quantum mechanical motions of the electrons around the nuclei in
two scattering molecules one would make the approximation of taking the 
scattering or exchange amplitude to be given simply by the 
overlap of the wave function for the two incomming molecules with that 
of the two molecules after the scattering. Such an overlap approximation 
for the scattering or exchange  process when it goes fast, would be 
completely analogous to the type of approximation which we use in 
our calculations of the Veneziano model amplitude in our fake scattering theory.

In a hih energy hadron colission the meeting of the two hadrons goes rather 
fast compared to the moving around of the constituents / partons 
in the hadrons. So the usual low transverse momentum type of hadronic 
colissions are not so far from the described case of a rather fast 
exchange between the molecules compared to the moving around of the 
consituents. In this sense one might speculate whether the 
rather fast passge of the hadrons could be described as being 
close to being fake in the sense that the genuine interaction between the 
consituents first shows up before or after the main hitting passage.

When a couple of partons really hit each other there is a fast interaction 
taking place between the constituents it would not be analogous to the fake 
process even in the short passge time.


\section{Unitarity}

{\bf A Major Achievement of Phantasy Hamiltonian Formulation is 
Unitarity of Time-deveopment Operator.
 }

If the theory has a formal /phantasy time development given by 
a Hamiltonian $H_{phantasy}$ then we have automatically that 
developping during some time interval wil result in 
a unitary operator development.

Essentially unitary S-matrix.

{\bf Perturbation Expansion in Coefficient on the ``Phantasy Hamiltonian'' 
$H_{phantasy}$}

Really the overall scale of the $H_{phantasy}$ is a matter of the 
time unit. In fact there is no time in the theory before we 
introduce the phantasy degrees of freedom and make them move.\\

\vspace{0.2cm}

Natural to make perturbation theory in the coefficient on 
$H_{phantasy}$.\\

\vspace{0.2 cm}

Then we get one shift in the topology or way of connection of the 
cyclic chains for each order in the perturbation. That corresponds to 
different topologies of string surface diagrams as describing 
unitarity corrections to the Veneziano model.


\section{Conclusion}

This article was truly inspired by our novel string field theory 
on which we by now have worked for long, and believe to have formulated 
a theory in which many strings can be present, so that it is really 
a string field theory, in terms of what we called objects, which is 
really pieces of strings taken for right and left movers seperately.

The remarkable fact turning out of this our old formalism was that the 
objects, meaning the bits making up the right and left mover degrees of freedom
turned out having zero Hamiltonian, zero timedevelopment.

As a pedagogical excercise to study such a system like our objects 
with zero Hamiltonian we started by considering a second quantized 
system of infinitely heavy non-relativistic particles they namely have 
vanishing Hamiltonian if they do not have any interaction: 

\begin{itemize}
\item We have put forward a very trivial second quantized theory
(of infinitely heavy non-relativsic particles identified as our earlier 
``objects'') and assumed for it a Hamiltonian that is zero as operator.
So no timedevelopment in this ``fundamental'' theory.
(It is this one which is the analogue of the theory of the objects from 
our Novel String Field Theory.)
\item We can only make it more interesting or adjustable by assuming 
something about the state of it. Say by a density matrix $\rho_{fundamental}$.
We use this option to assume that the particles (=``objects'') sit 
in (long) closed chains  (cyclic chains). 
\item We interpret each cyclic chain to describe an open string 
in a string theory.


\item We introduced a phantasy system of degrees of freedom by introducing
a ``successor function'' f, which puts all the ``objects'' ($\sim$ paricles )
into a series of closed chains, thereby making explicit that such chains 
are assumed to be present by the assumption about the likely state of the 
trivial second quantized system.

\item Mostly we imagine the cyclic ordering is given by the 
``fundamental'' state of the trival theory, but in some cases it will 
be ambiguous which chains there are. Then it is we introduce
the fake/phantasy/f-variable to distinguish possibilities. 


\item Then the idea was to make a Hamiltonian supposed to mainly 
make this fake degree of freedom move, but approximately to avoid 
varying the ``fundamantal'' degrees of freedom. 

With this we then get a quite phantasy time. We only get timedevelopment 
due to the phantasy degrees of freedom.  
 
This could be used to realize the philosophy that the very concept of time 
is indeed a fake concept, so that at the fundamental level there is no 
time, but only a static state of the universe. Then only by introduction of 
a fake 
overbuilding (analogous to our phantasy successor function $f$ we obtain 
a world seemingly having a time-concept.

Indeed different moments would corresponds to just different ways 
of looking at the very same state, whatever the moment in qustion.
\end{itemize}


{\bf Conclusion on Hopes and Applications}

\begin{itemize}
\item Really the formulation of ours is a {\bf solution} of second quantized 
string theory, in the sense that we could say we solved the time development 
by identifying string theory with several strings with a theory without 
time development.

\item Hope to generalize our ``object'' picture to different models 
which have the same great property as string theory of {\bf not 
having usual divergencies}! This would be absolutely needed in high 
dimensions, because with point particles high dimensions cause 
rather hopeless divergencies.
\item As a special case we may generalize to p-adic\cite{padic6, Branko} 
Veneziano model\cite{NN7}.
\end{itemize}


\section*{Acknowledgement}

One of us (H.B.N.) acknowledges the 
Niels Bohr Institute for allowing him to work 
as emeritus and for partial economic support.
Also thanks food etc. support from the Corfu conference
and to Norma Mankoc Borstnik for asking for a 
way to get meaningful quantum field theories in higher than 4 
dimensions. The thinking on hadronic like bound states could namely be looked upon 
as an attempt to find such a scheme using bound states as the theory behind the 
particles for which to make the convergent theory.

M. Ninomiya acknowledges Yukawa Institute of Theoretical Physics, 
Kyoto University, and also   the Niels Bohr Institute
and Niels Bohr International Academy for giving him
 very good hospitality
during his stay. M.N. also acknowledges at 
Yuji Sugawara Lab. 
Science and Engeneering, 
Department of physics sciences Ritsumeikan 
University, 
Kusatsu Campus for allowing him 
 as a visiting 
Researcher.

\end{document}